\input amstex.tex
\input amsppt.sty   
\vsize = 8.5 true in
\hsize=6.2 true in
\nologo
\NoRunningHeads        
\parskip=\medskipamount
        \lineskip=2pt\baselineskip=18pt\lineskiplimit=0pt
       
        \TagsOnRight
        \NoBlackBoxes

        \topmatter
        \title
        Approach to equilibrium for the stochastic NLS 
        \endtitle
       
\author
 J. L. Lebowitz, Ph. Mounaix and W.-M.~Wang 
\endauthor        

\address
Departments of Mathematics and Physics, Rutgers University, Piscataway, NJ 08854-8019,
\endaddress
\email  lebowitz\@math.rutgers.edu
\endemail

\address
Centre de Physique Th\'eorique, UMR 7644 CNRS, Ecole Polytechnique, 91128 Palaiseau Cedex
\endaddress
\email philippe.mounaix\@cpht.polytechnique.fr
\endemail

\address
CNRS and Department of Mathematics, Universit\'e Cergy-Pontoise, 95302 Cergy-Pontoise Cedex 
\endaddress
\email wei-min.wang\@math.cnrs.fr
\endemail

\abstract
We study the approach to equilibrium, described by a Gibbs measure, for a system on a 
$d$-dimensional torus evolving according to a stochastic nonlinear Schr\"odinger equation (SNLS)
with a high frequency truncation. We prove exponential 
approach to the truncated Gibbs measure both for the focusing and defocusing cases when the dynamics is constrained
via suitable boundary conditions to regions of the Fourier space where the Hamiltonian is convex. Our method is 
based on establishing a spectral gap for the non self-adjoint Fokker-Planck operator governing the time 
evolution of the measure, which is {\it uniform} in the frequency truncation $N$. The limit $N\to\infty$ is discussed.
\endabstract

        \bigskip\bigskip
        \bigskip
       \toc
       \bigskip
       \bigskip 
       \widestnumber
       \head {Table of Contents}
        \head 1. Introduction 
        \endhead
        \head 2. Truncated Fourier space formulation and the Theorem
        \endhead 
        \head 3. Sobolev embedding and convex Hamiltonian
        \endhead
        \head 4. The Witten-Hodge complex and spectral gap
        \endhead 
        \head 5. Exponential approach to equilibrium
        \endhead
        \head 6. The limit $N\to\infty$
        \endhead
         \head 7. Appendix: The Witten-Hodge complex
        \endhead
       \endtoc
        \endtopmatter
        \vfill\eject
        \bigskip
\document
\head{\bf 1. Introduction}\endhead
The nonlinear Schr\"odinger equation is used to model a variety of phenomena
in different physical systems, see e.g., \cite{LRS, MCL} and references therein. It comes in two versions: the focusing and the 
defocusing cases. These correspond respectively to the sign of $\lambda$ 
in front of the nonlinear term in the Hamiltonian $H$ of the (isolated) system,
$$H:=H( u, \bar u)=\int_{\Bbb T^d}( |\nabla u|^2+m|u|^2)dx+\frac{2\lambda} {p}\int_{\Bbb T^d} |u|^{p}dx, \tag 1.1$$
where $p$ is a positive even integer and $m\geq 0$. Here $u$ is a complex valued field (function) on the $d$-torus: 
$\Bbb T^d:=[0, 2\pi]^d$ after identifying the end points. 
The time evolution of $u$ satisfies the equation:
$$\aligned i\frac{\partial u}{\partial t}&=-\frac{\partial H}{\partial \bar u}\\
&=-\Delta u+mu+\lambda|u|^{p-2}u.\endaligned\tag 1.2$$
(Note that (1.1) differs from the $H$ in \cite{LRS, MCL} by a factor of $2$.) 

In the defocusing case, $\lambda$ is positive, $H$ is bounded from below and one then has global existence and uniqueness 
of smooth solutions in $d\leq 4$ for appropriate $p$; more precisely, for $p\leq 4$ in $d=4$, $p\leq 6$ in $d=3$ and all $p$
in $d\leq 2$. This is the case of interest in constructive
quantum field theory \cite{GJ}. 

The situation is very different for the focusing case, $\lambda<0$. In this case $H$ is not bounded from below and there 
exists a critical $p_c$, $p_c= 2+4/d$, such that for $p<p_c$, there are unique global smooth solutions;
for $p=p_c$ there are unique global smooth solutions with small $L^2$ norm.  For $p>p_c$ but $p\leq 2d/(d-2)$, one needs to have small 
$H^1$ data to have unique global solutions, cf. e.g., \cite{Bo3}.

This dichotomy carries over to the possible existence of a (non-truncated) Gibbs measure with density 
$$\mu=Z^{-1}\exp{[-\beta\int_{\Bbb T^d} \frac{2\lambda} {p}:|u(x)|^{p}:dx]},\tag 1.3$$
with respect to the free field Gaussian measure, where $\beta>0$ and $:\, :$ indicates Wick ordering, which is needed
for $d\geq 2$ and $Z^{-1}$ is a normalization factor. 

These measures exist for the defocusing case, $\lambda>0$ for all $p$ in $d=1$ (without Wick ordering) and for $p\leq 6$
in $d=2$, and $p\leq 4$ in $d=3$ \cite{GJ}. Furthermore, despite the fact that this measure is supported on  
rough functions, Bourgain showed that it is invariant under the dynamics given by (1.2) for $d\leq 2$ \cite{Bo1, 2}.
This means that the dynamics can be defined (after Wick ordering modification in $d=2$) on a set 
of full measure with respect to $\mu$. 

The focusing case is more delicate. Since $H$ is unbounded from below, it is obvious that the measure 
$\mu$ cannot exist without some restrictions on its domain. It was shown by Lebowitz, Rose and Speer
in \cite{LRS} that in $d=1$, the Gibbs measure exists for $p=4$ when restricted to $L^2$ balls and 
that it exists for $p= 6$ with the additional condition of small $|\lambda|$, cf. also \cite{McKV, Z}.
In $d=2$, Jaffe \cite{J}  showed that
the measure exists for $p=3$ for real $u$ when restricted to $L^2$ balls and after Wick ordering; while Brydges and Slade \cite{BS} showed
that  this does not work when $p=4$. 

In this paper, we shall deal with both the focusing and defocusing cases when the system is in contact with a heat reservoir of the 
Ornstein-Uhlenbeck type at reciprocal temperature $\beta$. This problem was studied in \cite{MCL}, 
whose formulation we shall follow here, cf. also \cite{LRS}.
The time evolution is then described by the stochastic nonlinear Langevin equation:

$$du+(\nu+i)(-\Delta+m+\lambda |u|^{p-2}) u\, dt=\sqrt {2\beta^{-1}\nu} dW,\tag 1.4$$
where $\nu\geq 0$ is the friction and $W$ is a complex Wiener process, i.e., 
white noise in space and time. We note that for the moment, equation (1.4) is formal as 
it involves a very rough $L^1$ force on the right side. 

To give a meaning to (1.4), we work
in Fourier space and restrict the number of Fourier modes $n$ to be finite and constrain the Hamiltonian 
so that it remains bounded from below and is locally convex. We enforce this bound on the evolution by 
imposing appropriate Neumann boundary conditions in Fourier space. The problem then reduces to a finite dimensional 
Markov process on a compact space, whose stationary measure is
given by a truncated Gibbs measure $\mu_N$. This is formulated in sect. 2. 
We note that for fixed $N$, exponential approach to $\mu_N$ without any lower bound on the rate, follows from general probabilistic considerations, see \cite{MCL}.  

The main new result of this paper is that the approach to $\mu_N$ is exponential with an exponent given by the size 
of the spectral gap for the non self-adjoint Fokker-Planck operator, which is {\it uniform} in the truncation $N$. 
This is proven in sects. 3-5. The proof is made possible by identifying the Fokker-Planck operator with a non self-adjoint Witten-Hodge Laplacian 
and a spectral analysis. 

We mention that a related Witten-Hodge Laplacian in the self-adjoint setting was introduced by Sj\"ostrand \cite{S1} 
to study correlation functions in equilibrium statistical mechanics and used by Wang \cite{W} to study the 
parabolic Anderson model. 

The question of what happens to the truncated measure $\mu_N$ and the dynamics in the limit as 
$N\to\infty$ is discussed briefly in sect. 6. It appears that in $d=1$, the limit will coincide with the $\mu$ given in
(1.3), whenever the latter exists. The situation in $d\geq 2$ is less clear since Wick ordering takes away convexity. 

 \bigskip
\noindent {\it Acknowledgement. }  W.-M.W. thanks J. Sj\"ostrand for several useful conversations. J.L. and W.-M.W. thank D. Brydges and G. Gallavotti for clarifications on the construction 
of the Gibbs measure. J.L. and Ph.M. thank P. Collet for many helpful discussions. J.L. is partially supported by the grants NSF DMR 08021220 and AFOSR AF-FA9550-10; W.-M.W. is partially supported by 
the grant ANR-10-JCJC0109 and the Qiushi professorial chair of mathematics at Zhejiang University. 
\smallskip
\head 2. {\bf Truncated Fourier space formulation and the Theorem}\endhead 
Let $n\in\Bbb Z^d$ be the dual variable to $x\in\Bbb T^d$
and $a=\{a_n\}$ the dual of $u$: 
$$u=\sum_{n\in\Bbb Z^d} e^{-in\cdot x}a_n.$$ 
We rewrite equation (1.3) using Fourier series. 
Let $\Phi$ be the corresponding sum in the Fourier space for $H$:
$$\aligned&\Phi(a)/(2\pi)^d:=\Phi(a,\bar a)/(2\pi)^d\\
&=\sum_{n\in\Bbb Z^d}(n^2+1)|a_n|^2+\frac{2\lambda}{p}\sum_{j_1-j_2...+j_{p-1}-j_{p}=0} a_{j_1}\bar a_{j_2}... a_{j_{p-1}}\bar a_{j_{p}},\endaligned\tag 2.1$$ 
where $a_n\in\Bbb C$, $n^2$ stands for $|n|^2$ and for simplicity we have set $m=1$. 
We now restrict to $u$ such that $a_n=0$ for $|n|>N$ and make estimates {\it uniform} in $N$, cf. the remark above (5.1) in sect. 5.
Since all estimates are uniform, we will generally omit $N$ as a subscript,
except in the statements of the Theorem and its Corollary. For notational simplicity, we set $\beta=2$ and rescale time by $(2\pi)^d$.
 
We have the following equation for $a$:   
$$\dot a_n=-i\frac{\partial \Phi}{\partial \bar a_n}-\nu \frac{\partial \Phi}{\partial \bar a_n}+\sqrt {\nu} \Gamma_n,\qquad |n|\leq N, \tag 2.2$$
where the $\Gamma_n$ are independent centered complex Gaussian white noises with 
$$\langle \Gamma_n(t)\bar \Gamma_n(t')\rangle=2\delta(t-t').\tag 2.3$$ 

The Fokker - Planck equation describes the time evolution of the probability density
$\tilde P(a, t)$ with initial distribution $\tilde P(a, 0)$ for a process given by a stochastic PDE.
Conjugating by $e^{\pm \Phi}$ and setting $P=e^{\Phi}\tilde P$,  the Fokker - Planck equation for (2.2) has the form: 
$$\partial_t P(a, t)+LP(a, t)=0,\tag 2.4$$
where 
$$L=-\nu\sum_n\frac{\partial^2}{\partial \bar a_n\partial a_n}+\nu\Vert\nabla\Phi\Vert^2-\nu\Delta\Phi+\Cal H_{\Phi},\quad{\nu>0}.\tag 2.5$$
Here $$\nabla\Phi \text{ is the vector }\{\partial_{\bar a_n} \Phi, \partial_{a_{n}} \Phi\},\tag 2.6$$
$$\Delta \Phi=\sum_n\frac{\partial^2}{\partial \bar a_n\partial a_n}\Phi,\tag 2.7$$ and 
$$\Cal H_\Phi=2\sum_n (\partial_{a_n}\Phi\partial_{\bar a_n}- \partial_{\bar a_n}\Phi\partial_{a_n})\tag 2.8$$
is the Hamiltonian vector field for the (finite dimensional) Hamiltonian $\Phi$
and is {\it anti} self-adjoint. (All indices above are restricted to $|n|\leq N$). Thus the Fokker - Planck operator $L$ is {\it non} self-adjoint.
This is a distinguishing feature of the present construction.  

In order for the Gibbs measure $\mu_N\sim e^{-2\Phi_N(a)}$ to be stationary for the Fokker-Planck equation (2.4) in the focusing case, we need to 
restrict to appropriate domains in $a$.
Toward that end, we define the $H^s$ balls and the Hamiltonian $\Bbb H$-ball of radius $\sqrt B$ to be the set of $a$ such that, respectively 
$$\sum_{n\in\Bbb Z^d} (n^{2s}+1)|a_n|^2<B,\quad B>0\tag 2.9$$
and
$$|\Phi(a)|/(2\pi)^d<B, \quad B>0,\tag 2.10$$
where $\Phi(a)$ is defined in (2.1). 

Since we are considering $u$ such that $a_n=0$ for $|n|>N$, the sums in (2.9, 2.10) are restricted to $|n|\leq N$. We take as our domain the Hamiltonian $\Bbb H$-balls and impose adapted 
Neumann boundary conditions on $\partial \Bbb H$ -- the energy surface: 
 $$\eta\cdot (-\nu\nabla\Phi-\nu\nabla+h_\Phi)P|_{\partial \Bbb H}=0,\tag 2.11$$
where $\eta$ is the normal direction, assuming that it is well-defined and 
$$h_\Phi=\{2\partial_{a_n}\Phi, -2\partial_{\bar a_n}\Phi\}$$
is a vector, displayed using the same convention as in (2.6).  On occasions, we will also call $h_\Phi$ 
the Hamiltonian vector field. It is important 
to remark that since $h_\Phi$ is tangential, the above boundary
condition is equivalent to 
$$\eta\cdot (-\nu\nabla\Phi-\nu\nabla)P|_{\partial \Bbb H}=0,\tag 2.12$$
which plays a pivotal role in establishing a spectral gap {\it uniform} in $N$. 

In the focusing case, the $\Bbb H$-ball is not necessarily a connected set. The theorem
below addresses the component that contains the origin. We call it 
the $\Bbb H_0$-ball.  In the defocusing case, $\Bbb H_0$ coincides with $\Bbb H$.

Our main results are : 

\proclaim {Theorem} For $d=1$, $2$, $p\in 2\Bbb N$ arbitrary, $d=3$, $p=4$, $6$ and $d=4$, $p=4$, there exist $C$, $c>0$ such that uniformly in $N$, the Fokker-Planck semi-group satisfies
$$\Vert e^{-tL_N}-\Pi_0\Vert_{\Cal L(L^2(\Bbb H_0), L^2(\Bbb H_0))}\leq Ce^{-ct},\quad t\geq 0$$
for sufficiently small $|\lambda|$, 
in a given Hamiltonian $\Bbb H_0$-balls with the Neumann boundary condition (2.11)
on the Hamiltonian surface $\partial \Bbb H_0$, where $\Pi_0$ is the projection onto the unique ground state of $L_N$:
$$\psi_N=C_Ne^{-\Phi_N},$$ and $C_N$ is such that the truncated Gibbs measure $\mu_N:=\psi_N^2$ is normalized: $$\int_{\Bbb H_0} \psi_N^2\prod_{|n|\leq N}da_nd\bar a_n=1,$$
and we have indicated explicitly the truncation $N$ on the $\Phi(a)$ defined in (2.1).
\endproclaim

\proclaim {Corollary} In the defocusing case, the Theorem holds for all $\lambda\geq 0$ under the same conditions on $d$ and $p$.
\endproclaim

We note that in the $1$d focusing case,  for $p=4$ and $6$, the Theorem remains valid with the Hamiltonian $\Bbb H_0$-ball
replaced by an $\Bbb L^2$-ball, cf. sect. 6. This is because the Hamiltonian vector field $h_\Phi$ is also tangential to the $\Bbb L^2$-sphere
and (2.12) remains valid on the sphere. Without dissipation and forcing, this Hamiltonian geometry leads to the conservation of mass, i.e., the $L^2$ norm and energy, essential
for the global existence and uniqueness of solutions for subcritical (or critical) dispersive NLS. With dissipation and forcing, the conservation laws
are no longer there, but it is important to recognize that the {\it geometry} remains and is precisely what enables the present construction.
The $1$d focusing case has particular interest as the limiting measure $\mu$ is known to exist \cite{LRS}, cf. sect. 6.

Previous related results on stochastic NLS, e.g.  \cite{EH, KuS, O} pertain either to the focusing $L^2$- subcritical case, namely $p\leq 4$, the $1$d focusing case, 
where there are no blow-up solutions, or to the defocusing case. The forces there are smooth in the spatial variable $x$. So one works directly with the 
(non-truncated) SNLS. Generally speaking, this approach does not yield the explicit form of the invariant measure. For results in the purely dissipative case, cf. \cite{DD}.
\smallskip
Equation (1.3) balances dissipation with a rough force. The main novelty of the
present construction is to work directly in the phase - space (the cotangent bundle) using the Fokker - Planck operator and impose boundary
conditions there. This enables us to smooth the rough force by convexity of the Hamiltonian and treat the critical cases.
\bigskip
\head{\bf 3. Sobolev embedding and convex Hamiltonian}\endhead
We establish convexity of $\Phi$ uniformly in $N$ and prove
\proclaim{Proposition} Under the same conditions on $d$ and $p$ as in the Theorem,  the Hessian of the truncated $\Phi$ satisfies
$$c\Bbb I<\text {Hess } \Phi:=\Phi''<C\Bbb I,\tag 3.1$$ 
in the $H^1$ ball defined in (2.9), where $c, C>0$ are constants which do not depend on $N$, if $|\lambda|$ is 
sufficiently small.
Here $\text {Hess } \Phi$ is considered as an operator on $\ell^2 ([-N, N]^d)\times \ell^2 ([-N, N]^d)$.
If $\Phi$ is defocusing, then $\lambda$ can be taken to be $1$ and $C=C(B)$.
\endproclaim

\demo{Proof}
We write the matrix operator $\Phi''$ in the form  
$\Phi''=M_1+M_2$, where 
$$\aligned &M_1=\pmatrix [[\partial_{\bar a_j}\partial _{ a_k}\Phi]]&0\\0&[[\partial_{a_j}\partial_ {\bar a_k}\Phi]]\endpmatrix,\\
&M_2= \pmatrix 0&[[\partial _{\bar a_j}\partial_{\bar a_k}\Phi]]\\[[\partial_ {a_j}\partial_{a_k}\Phi]]&0\endpmatrix,\endaligned\tag 3.2$$
and $[[\quad]]$ denotes the matrix of second order partial derivatives.

Now to prove (3.1), it suffices to prove that 
$$(v, \Phi'' v)\geq c (v,v)\tag 3.3$$ for all $v=\{w_j, \bar w_j\}\in\ell^2\times\ell^2$. 
The quantity $(v, \Phi'' v)$ is a sum of terms of the form
$M_{1,\, jk}w_j\bar w_k$ or $M_{2,\, jk}w_jw_k$ and their complex conjugates.
Let $W$ be the function with Fourier coefficients $w_j$ and $u$ the function with Fourier coefficients $a_j$ ($|j|\leq N$).
We observe that since $\Phi$ is a sum of homogeneous polynomials in $a_j, \bar a_k$, the terms that occur in the 
sum are exactly the same as the terms in the evaluation of the following integrals:
$$\Vert W\Vert^2_{H^1}:=\Vert \nabla W\Vert^2_{L^2}+\Vert W\Vert^2_{L^2}=\sum (n^2+1)|w_n|^2,\tag 3.4$$
$$\Vert u^{(p/2-1)} W\Vert^2_{L^2},\, \int W^2\bar u^2|u|^{(p-4)}dx\quad\text{and its complex conjugate},\tag 3.5$$
where here and below $\int \, \cdot \, \,dx$ denotes integration over the torus $\Bbb T^d$.  

More precisely, 
$$(v, \Phi'' v)=2\Vert W\Vert^2_{H^1}+\lambda p\Vert u^{(p/2-1)} W\Vert^2_{L^2}+\lambda (p-2)\text{Re }\int W^2\bar u^2|u|^{p-4}dx.\tag 3.6$$
Using now the inequalities 
$$\frac{2}{p}\Vert u^{(p/2-1)} W\Vert^2_{L^2}\leq\Vert u^{(p/2-1)} W\Vert^2_{L^2}+(1-2/p)\text{Re }\int W^2\bar u^2|u|^{p-4}dx\leq 2 \Vert u^{(p/2-1)} W\Vert^2_{L^2},\tag 3.7$$
we have in the focusing case,
$$2\Vert W\Vert^2_{H^1}-2|\lambda| p\Vert u^{(p/2-1)} W\Vert^2_{L^2}\leq (v, \Phi'' v)\leq 2\Vert W\Vert^2_{H^1}+2|\lambda|p\Vert u^{(p/2-1)} W\Vert^2_{L^2}\tag 3.8$$
and in the defocusing case
$$2\Vert W\Vert^2_{H^1}\leq (v, \Phi'' v)\leq 2\Vert W\Vert^2_{H^1}+2|\lambda| p\Vert u^{(p/2-1)} W\Vert^2_{L^2}.\tag 3.9$$
Standard Sobolev embedding on $\Bbb T^d$, namely $$\Vert \cdot\Vert_{L^{p}}\leq C_{d, p}\Vert\cdot\Vert _{H^{d/2(1-2/p)}}\tag 3.10$$ then gives
under the same conditions as in the Theorem on $d$ and $p$ that 
$$\Vert u^{(p/2-1)} W\Vert^2_{L^2}\leq C_{d, p}\Vert u\Vert_{H^1}^{p-2}\Vert W\Vert_{H^1}^2.\tag 3.11$$
Using (3.11) in (3.8, 3.9) and also smallness of $|\lambda|$ in (3.8), proves 
the proposition. 
\hfill $\square$
\enddemo
\smallskip
\head{\bf 4. The Witten-Hodge Laplacian and spectral gap}\endhead
We now prove that the Fokker-Planck operator $L$  in (2.5) 
restricted to Hamiltonian $\Bbb H$-balls defined in (2.9) has a spectral 
gap uniform in $N$ when the Hamiltonian is convex. 
We use the Witten-Hodge Laplacian formulation introduced in a related context in \cite{S1}.  We summarize 
some of the basic notions in the Appendix using the self-adjoint setting. Here we show that it corresponds
to the operator $L$ and gives the desired spectral gap. 
\smallskip
\noindent{\it The non self-adjoint Fokker-Planck operator}

Let $d$ be the exterior differentiation 
$$d:=\sum_n\partial_{b_n}db_n^{\wedge},$$
where $b_n$ stands for $a_n$ and $\bar a_n$ and  
$$d_{\Phi}:=e^{-\Phi}d e^{\Phi}=\sum_n(\partial_{b_n}+\partial_{b_n}\Phi )db_n^{\wedge}.\tag 4.1$$ 
We note that here we work on $\Bbb C^\Lambda$ with $\Lambda= [-N, N]^d$, which could be identified
with $(\Bbb R^2)^\Lambda\sim\Bbb R^{2|\Lambda|}$.

Let $$A=\pmatrix \nu\Bbb I &\Bbb I\\-\Bbb I&\nu\Bbb I\endpmatrix,$$
where the notation is the same as in (3.2) with each bloc being one 
of the four possible sectors: $\bar u u, \bar u\bar u, uu, u\bar u$. Define
the (formal) adjoint of $d_\Phi$ with respect to $A$ to be:
$$d_\Phi^{*, A}= \sum_n(-\partial_{\bar b_n}+\partial_{\bar b_n}\Phi )\circ A d\bar b_n^{\rfloor}.$$ 
(Properly speaking, $A$ is a map from the cotangent space ${(\Bbb C^{{[-N, N]^d}})}^*$ to the tangent space
$\Bbb C^{{[-N, N]}^d}$.)
The Witten-Hodge Laplacian is then defined as 
$$-\Delta_{\Phi, A}=d_\Phi^{*, A}d_\Phi+d_\Phi d_\Phi^{*, A},$$
cf. the paper of Bismut \cite{Bi} for the general construction and also the Appendix in sect. 7.

The reason that we introduce the Witten-Hodge Laplacian is that
when restricting to scalar functions, the $0$-forms, it is precisely the Fokker - Planck operator $L$ in (2.5) and we have the following identities:
$$-\Delta^{(0)}_{\Phi, A}=d_\Phi^{*, A}d_\Phi=d_\Phi^*\circ A\circ d_\Phi=L,\tag 4.2$$
where 
$$d_\Phi^{*}= \sum_n(-\partial_{\bar b_n}+\partial_{\bar b_n}\Phi )d\bar b_n^{\rfloor}.$$
We will also need it on $1$-forms: 
$$-\Delta^{(1)}_{\Phi, A}=-\Delta^{(0)}_{\Phi, A}\otimes\Bbb I+2\Phi''\circ A^t, \tag 4.3$$
cf., \cite{HHS, S2}. 

\noindent{\it Remark.} This is occasionally dubbed the supersymmetric approach, cf. e.g. \cite{TT-NK} 
for the physics literature. 

Let the Hamiltonian $\Bbb H$-ball be as defined in (2.10) and as before, $\Bbb H_0$-ball the connected component containing
the origin. The following basic spectral characterization of the Fokker -Planck operator follows readily from (4.2).
\proclaim{Lemma 1} Assume $\Bbb H_0$ is a convex set. 
The spectrum of $L$ in this ball with the Neumann boundary condition (2.11)  is contained in the sector $$\{\kappa\in\Bbb C|{\text Re } \kappa\geq \nu|{\text Im }\kappa|\}, 
\tag 4.4$$ and $0$ is a simple eigenvalue.
\endproclaim
\demo{Proof}
The spectrum of $L$, $\sigma(L)$ is contained in the numerical range of $L$: $$\{(u, Lu)\in\Bbb C|u \text{ in the domain of } L\}.$$ From (4.2)
for all $u$ in the domain of $L$ satisfying the Neumann boundary condition (2.11),
$$(u, Lu)=(u, -\Delta^{(0)}_{\Phi, A} u)=\nu\int_{\Bbb H_0} |d_{\Phi} u|^2dad\bar a+i\text { Im }\int_{\Bbb H_0} (d_{\Phi} u)^2dad\bar a,\tag 4.5$$
where $(d_{\Phi} u)^2$ denotes the sum of the square of components, with each component
defined as in (4.1).
Equation (4.5)  implies (4.4), since (4.5) equals zero if and only if $d_\Phi u=0$, this means $u$ is a multiple
of $e^{-\Phi}$, which concludes the proof.
\hfill $\square$
\enddemo 
Using the Witten-Hodge Laplacian (4.3) on $1$-forms, we have, moreover, the following spectral gap lemma, 
essential for the proof of the Theorem.

\proclaim{Lemma 2} Assume that in the $\Bbb H_0$-ball, $\Phi''$ satisfies $cI<\Phi''<CI$ for some $c, C>0$. Then for some $c_0=c_0(c)>0$, the Fokker - Planck operator $L$ with Neumann boundary 
conditions has the following properties uniformly in $N$: 

\item{(i)} $\{\sigma(L)\backslash\{0\}\}\cap \{z\in\Bbb C|\text{Re } z<c_0\}=\emptyset$;  
\item{(ii)} (up to constants) $e^{-\Phi}$ is the unique ground state with eigenvalue $0$.

Let $\Pi_0$ be the projection onto the normalized ground state. Then 
\item{(iii)} $\Vert (L-z)u\Vert\geq (c_0-\text{Re } z)\Vert u\Vert$ if $u=(1-\Pi_0) u$.
\endproclaim

\demo{Proof}
Assume $u$ is an eigenfunction of the Fokker -Planck operator $L$ with Neumann boundary conditions with eigenvalue $\kappa\neq 0$, i.e.,
$$Lu=-\Delta^{(0)}_{\Phi, A} u=\kappa u, \quad \kappa\neq 0.\tag 4.6$$
Then from Lemma 1, $d_\Phi u\neq 0$. Operating on equation (4.6) using $d_\Phi$ and taking the scalar product with $d_\Phi u$ , we have
$$(d_\Phi u, [d_\Phi d_\Phi^{*, A}]d_\Phi u)=\kappa(d_\Phi u,  d_\Phi u).$$  
Writing $v$ for $d_\Phi u$, we have equivalently
$$(v, -\Delta^{(1)}_{\Phi, A}v)=(v, [-\Delta^{(0)}_{\Phi, A}\otimes\Bbb I+2\Phi''\circ A^t]v)=\kappa (v, v), \tag 4.7$$
where we used (4.2). 

Taking the real part of (4.6), we obtain 
$$\text { Re } \kappa \Vert v\Vert^2=(v,  [-\Delta^{(0)}_{\Phi}\otimes\Bbb I] v)+ 2\nu(v, \Phi'' v),$$
where $-\Delta ^{(0)}_{\Phi}=d^*_\Phi d_{\Phi}$ is the self-adjoint Laplacian on $0$-forms. 

Using the convexity of $\Phi$,
we then obtain (i) with $c_0=2c\nu>0$ uniformly in $N$. Here we also used the fact that the self-adjoint 
Laplacian $-\Delta ^{(0)}_{\Phi}$ has the {\it same} Neumann boundary condition as $-\Delta ^{(0)}_{\Phi, A}$.
We remark that in fact stronger results are known under appropriate conditions, namely 
$$\sigma (-\Delta^{(0)}_{\Phi, A})\backslash{\{0\}}\subset \sigma (-\Delta^{(1)}_{\Phi, A}),$$
cf. \cite{HHS}.

(ii) follows from (i) and Lemma 1. To prove (iii), we write
$$\aligned \Vert &(L-z)u\Vert\Vert u\Vert=\Vert (-\Delta^{(0)}_{\Phi, A}-z)u\Vert\Vert u\Vert\geq | (-\Delta^{(0)}_{\Phi, A}-z)u, u)|\\
&\geq  | \text { Re }((-\Delta^{(0)}_{\Phi, A}-z)u, u)|\\
&=|((-\Delta^{(0)}_{\Phi}-\text{ Re }z) u, u)|\\
&\geq (c_0-\text { Re } z)(u, u)\endaligned\tag 4.8$$
for all $u=(1-\Pi_0)u$. 

Here we used self-adjointness of $-\Delta^{(0)}_{\Phi}$, the fact that up to constants $e^{-\Phi}$ is also the unique ground state of  $-\Delta^{(0)}_{\Phi}$,
i.e., 
$$\Pi_0(-\Delta^{(0)}_{\Phi})=\Pi_0(-\Delta^{(0)}_{\Phi, A})$$
and eigenfunction (of $-\Delta^{(0)}_{\Phi}$) expansion of $u$.
\hfill $\square$
\enddemo

\noindent{\it Remark.} Since $L$ is non self-adjoint, it is no longer true that the resolvent 
at $z$ is bounded above by the inverse of the distance of $z$ to the spectrum. This is because of
the non-commutativity of the self-adjoint and anti self-adjoint components and hence the 
necessity of the type of arguments in (4.8).

\bigskip
\head{\bf 5. Exponential approach to equilibrium}\endhead
\demo{Proof of the Theorem}
We only need to verify the conditions in Lemmas 1 and 2, namely convexity of the Hamiltonian in $\Bbb H_0$. 
The rest will follow by applying these two lemmas and contour integration. Recall that the $H^s$ and the
Hamiltonian $\Bbb H$-balls are as defined in (2.9, 2.10) and we restrict to functions $u$ such that
its Fourier coefficients $a_n=0$ for $|n|>N$. So the sums below are to be understood as:
$$\sum:=\sum_{n\in\Bbb Z^d}=\sum_{|n|\leq N};$$
in other words, the $B$ below is {\it fixed} independent of  $N$. Therefore the radius of the various ``balls" are fixed.
It is only the dimension of the balls that varies with $N$. 

\noindent{\it Remark.} For most of the applications, $u=\pi \tilde u$, where $\pi$ is the projection 
onto the first ``$N$" Fourier modes while $\tilde u$ has full Fourier support and is in at least one of
 the ``balls".

Define the Hamiltonian $\Bbb H$-ball as before: 
$$\aligned |\Phi(a)/(2\pi)^d|=&|\sum(n^2+1)|a_n|^2+\frac{2\lambda}{p}\sum_{j_1-j_2...+j_{p-1}-j_{p}=0} a_{j_1}\bar a_{j_2}... a_{j_{p-1}}\bar a_{j_{p}}|\\
&<B.\endaligned\tag 5.1$$
Using the results of sect. 3, we have that
if $a$ is in the above $\Bbb H$-ball, then for the defocusing case, $\lambda>0$, $\Phi$ is convex and $(v,\Phi'' v)<C$.
So $\Bbb H$ is a convex set with the well-defined boundary $$\partial \Bbb H=\{a|\,\Phi(a)/(2\pi)^d=B\},$$
and $$\sum(n^2+1)|a_n|^2<B$$
for $a$ in $\Bbb H$.
 
For the focusing case, $\lambda<0$, assume we look at the connected component containing the origin, namely $\Bbb H_0\ni 0$. 
Let $A$ be the set 
$$A=\{a|\sum(n^2+1)|a_n|^2<5B\}, \tag 5.2$$ 
and
$$\Bbb H_A:=\Bbb H_0\cap A.$$
The Proposition then gives that $\Phi$ is convex in $A$ and moreover $\Phi(\cdot)$ is equivalent 
to the $H^1$ norm:
$$\frac{1}{2}\Vert f\Vert_{H_1}<(1-C|\lambda|B^p)\Vert f\Vert_{H_1}<\sqrt {\frac{\Phi (f)}{(2\pi)^d}}<(1+C|\lambda|B^p)\Vert f\Vert_{H_1}<2\Vert f\Vert_{H_1},\tag 5.3$$
for $|\lambda|\ll 1$ and where $f$ is the function with Fourier coefficients $a\in A$. 

Equation (5.3) together with (5.2) give that $\Bbb H_0$ is strictly contained in $A$: 
$$\Bbb H_A\equiv \Bbb H_0.$$
So $\Bbb H_0$ is a convex set
with the well-defined boundary $$\partial \Bbb H_0=\{a|\,\Phi(a)/(2\pi)^d=B\}\subset A.$$

We now evaluate the semi-group $e^{-tL}$ using the contour $\Gamma=\Gamma_1+\Gamma_2$,
where $\Gamma_1$ is compact enclosing $0$ and $\text {Re }\Gamma_1\leq c'<c_0$, the $c_0$ in
Lemma 2; and $\Gamma_2$ is defined by
$$\{z|\,|\text{Im }z|=|\text{Re }z|^2+D,\text{ if } \text{ Re }z>c' \text{ and } |\text{Im }z|\leq {c'}^2+D \text{ if } \text{ Re }z=c' \text{ for some } D>0\},$$
so that 
$$\aligned e^{-tL}&=\frac{1}{2\pi i}\oint_\Gamma {e^{-tz}}(z-L)^{-1}dz\\
&=\frac{1}{2\pi i}\oint_{\Gamma_1} {e^{-tz}}(z-L)^{-1}dz+\frac{1}{2\pi i}\oint_{\Gamma_2} {e^{-tz}}(z-L)^{-1}dz.\endaligned$$

It follows from Lemmas 1 and 2 that the above contours lie in the resolvent set and the integrals are well-defined. (i, ii) of Lemma 2 proceed to 
give that the first term is $\Pi_0$ and (iii) gives the exponential estimate for the second term, uniformly in $N$ and concludes
the proof. 
\hfill $\square$
\enddemo

\demo{Proof of Corollary}
This follows from global convexity of $\Phi$, which is moreover {\it uniform} in the radius $\sqrt B$ of the $\Bbb H$-ball,
since $H^1$ ball is equivalent to $\Bbb H$-ball in the defocusing case. 
Since
$$d\Phi=\sum_n\frac{\partial \Phi}{\partial b_n}db_n,$$
where $b_n=a_n$ and $\bar a_n$, let $u$ be the function with Fourier coefficients $\{a_n\}$, we have 
that $$\aligned 0<\sum |\frac{\partial \Phi}{\partial a_n}|^2&\leq 4\pi^{2d}(\Vert u\Vert_{H^2}+|\lambda|\Vert u\Vert^{p-1}_{L^{2p-2}})^2\\
&< 4\pi^{2d}(\Vert u\Vert_{H^2}+|\lambda|C_{d, p}\Vert u\Vert^{p-1}_{H^2})^2<\infty,\endaligned$$
where we used convexity for the lower bound, (3.10) and the restrictions on $d$ and $p$. So the energy surfaces $\partial\Bbb H$ is  well-defined
if $u\in H^1\cap H^2$. Using the density of $H^2$ in $H^1$, we reach the conclusion of the Corollary. 
\hfill $\square$  
\enddemo

\bigskip
\head{\bf 6. The limit $N\to\infty$}\endhead
As discussed in the introduction, the existence of a Gibbs measure $\mu\sim \exp[-\beta H]$ for the $H$ 
in (2.1) is a problem which has been studied extensively. It corresponds in $d=1$ to the problem of the behavior of the limit $N\to\infty$
of the measure  $\mu_N\sim \exp[-\beta \Phi_N]$. These measures exist for all $p$ in the defocusing case and 
for $p\leq 6$ for the focusing case under some restrictions. In $d\geq 2$, even the defocusing case requires 
Wick ordering, see (1.3), and it does not exist for the focusing case.

In $d=2$, the Hamiltonian dynamics corresponding to $\nu=0$, have been modified by Bourgain to include Wick ordering and shown,
as mentioned earlier, to be well defined for the defocusing case when $p=4$.  Whether one can make a similar 
modification to the Langevin dynamics, $\nu>0$, is an open and intriguing question. We note however that 
the Wick ordered nonlinear term  $:u^p:$ is not convex. So the argument in this paper leading to the existence of a spectral gap
would not hold for the modified dynamics. Of course the result might still be true. But this is another question.

The question of what happens to the dynamics generated by (2.2) or the Fokker - Planck equation (2.4)  in the 
limit $N\to\infty$ is therefore of relevance here primarily in the case $d=1$. We consider the defocusing and the focusing
cases below. 

 \noindent{\it The defocusing SNLS}

In this case the estimates in the Corollary are not only uniform in $N$, but also in the radius $\sqrt B$ of the Hamiltonian
ball, which is equivalent to the $H^1$ norm. For $d=1$, letting $B\to\infty$ and subsequently $N\to\infty$, should therefore lead to the limiting measure with density: 
$$\mu=Z^{-1}\exp{[-\frac{\beta\lambda} {p}\int_{\Bbb T}|u(x)|^{p}dx]},$$
with respect to the free field Gaussian measure. A full proof is under investigation.

\noindent{\it The $1d$ focusing SNLS}

In the $1d$ focusing case, the limiting measures when $p=4$ and for small $|\lambda|$ also $p=6$ are known to exist \cite{LRS} and are invariant \cite{Bo1} under the corresponding 
dispersive NLS dynamics, i.e., (1.3) when $\nu=0$.  When $\nu>0$, the Theorem indicates a spectral gap {\it uniform} in $N$, albeit restricted to a Hamiltonian
ball in which it is convex. 

For fixed $N$, this restriction to a Hamiltonian ball can be replaced by a restriction to an $L^2$ ball with Neumann boundary conditions, as mentioned after the Theorem in sect. 2, since the Hamiltonian vector field is also tangential to the $L^2$ sphere. We then have that for both $p=4$ and $6$, the Hamiltonian is convex (uniformly in $N$) for small $|\lambda|$ depending on the radius of the $L^2$ ball using Sobolev embedding and interpolation.
In the focusing case, the Hessian of the Hamiltonian is bounded above by a constant (which only depends on $N$).  So using the same arguments as in sects. 4 and 5, the Fokker-Planck operator has a spectral gap uniform in $N$. 

This $L^2$ Neumann restriction is a priori compatible with the ensuing $N\to\infty$ limit as the Brownian motion in $d=1$ is supported in $H^{1/2^-}\subset L^2$
and should lead to the limiting measure constructed in \cite{LRS}. The question would then concern its invariance with respect to the SNLS dynamics in (1.3).
The general procedure could be akin to that in \cite{Bo1} on the corresponding Hamiltonian dynamics. Alternatively one could try to consider the limit $N\to\infty$ 
of the equation (2.4). See \cite{H} for some results in a related but finite $N$ setting. 
\bigskip
\head{\bf 7. Appendix: The Witten-Hodge complex}\endhead

Below we make a short introduction to the Witten-Hodge complex (originally introduced in \cite {Wi})
using the self-adjoint setting. This is because the structure remains the same and the notations are 
a bit simpler, cf. Chap. 11, in particular Chap. 11.4 in \cite{CFKS}. 

Toward that purpose, let $\phi\in C^\infty(\Bbb R^{N}; \Bbb R)$.
Let $d$ be the usual exterior differentiation:
$$d=\sum^N_{j=1}\partial_{x_{j}}dx_j^{\wedge}$$
 and 
$$d_{\phi}=e^{-\phi}  d e^{\phi}=d+d\phi^{\wedge}=\sum_{j\in\Lambda}z_jdx_j^{\wedge},\tag
7.1$$ 
where
$$z_j=\frac {\partial}{\partial x_j}+\frac
{\partial\phi}{\partial x_j}.\tag 7.2$$ (For the calculus of differential forms, see for example \cite{Sp}.)

If $f$ is a form of degree $m$, then $d_\phi f$ is a form of
degree $m+1$. For example, if $f$ is a $0$-form, i.e., a scalar function in $C^\infty(\Bbb R^N; \Bbb R)$, 
then
$$d_\phi f=\sum_{j=1}^N (z_j f) dx_j$$
is a $1$-form, which we may identify with a vector valued function $F$ with the components:
$$F_j(x)=(z_j f) (x),$$
i.e., a function in $C^\infty(\Bbb R^{N}; \Bbb R^N)$.
(We note that when $\phi=0$, $d_\phi f=df$, which is just the usual differential of $f$.)
If $f$ is a $1$-form, $f=\sum_j f_j dx_j$ then 
$$d_\phi f=\sum_{i<j} (z_if_j) dx_i\wedge dx_j$$
is a $2$-form, which we may identify with an $N\times N$ antisymmetric matrix function $M$ with the entries:
$$M_{ij}(x)=-M_{ji}(x)=(z_if_j) (x),$$ i.e., a function in $C^\infty(\Bbb R^{N}; \Bbb R^N\wedge\Bbb R^N)$.
The operator $z_j$ can be seen as an annihilation  operator in view of its action on $e^{-\phi}$, namely 
$$z_j e^{-\phi}=0, \text{ for } j=1, 2, ...,N. $$ 

On the space of $m$-forms, one may define an $L^2$ scalar product
for two $m$-forms $\omega$ and $v$: $(\omega, v)$.
Specializing to $v=d_\phi f$,
where $f$ is an $(m-1)$- form, we define the adjoint operator
$d^{*}_{\phi}$ of $d_{\phi}$ so that:
$$(\omega, d_{\phi}f)=(d^{*}_{\phi}\omega, f).$$
This gives the formal adjoint to be 
$$d^{*}_{\phi}=e^{\phi} 
d^{*}e^{-\phi}=\sum^N_{j=1}z^{*}_jdx_j^{\rfloor},\tag 7.3$$ where 
$$z^{*}_j=-\frac {\partial}{\partial x_j}+\frac {\partial\phi}{\partial x_j}\tag
7.4$$ 
and $\rfloor$ is the usual contraction which lowers the degree of the forms.

If $f$ is a form of degree $m$, then $d^*_\phi f$ is a form of
degree $m-1$. For example, if $$f=\sum^N_{j=1} f_jdx_j$$ 
is a $1$-form, then 
$$d^*_\phi f=\sum_{j=1}^N z^*_j f_j$$ 
is a $0$-form, i.e., a scalar function in $C^\infty(\Bbb R^{N}; \Bbb R)$.
(We note that when $\phi=0$, $d^*_\phi f=d^*f$ is just the
usual divergence of $f$.) If $f$ is a $0$-form, then $d^*_\phi f=0$.

The operator $z^{*}_j$ can be viewed as a creation operator.
For example, when $N=1$ and $\phi=x^2$,  $z^{*}:=z^{*}_j$ generates the
first Hermite polynomial. We  have the commutation relation: 
$$[z_j,z_k^{*}]=2\partial_j\partial_k\phi,\tag 7.5$$ which plays an  important
role. We check easily that indeed $$d_\phi 
d_\phi=d^{*}_{\phi}d^{*}_{\phi}=0.\tag 7.6$$ 

Using $d_\phi$, 
${d^*_\phi}$, we define the Witten Laplacian, 
$$-\Delta_{\phi}=d^{*}_{\phi}{d_\phi}+{d_\phi}d^{*}_{\phi},\tag 7.7$$ on
$C^{\infty}(\Bbb R^{N};\wedge^\ell\Bbb R^{N})$ ($  1\leq\ell\leq
N$), where $\wedge^\ell\Bbb R^{N}$ denotes the $\ell^{\text{th}}$ anti-symmetric tensor 
product of $\Bbb R^{N}$
with $\wedge\Bbb R^N:=\Bbb R^N$, $\wedge^2\Bbb R^N:=\Bbb R^N\wedge\Bbb R^N$ etc.
For example, if $\ell=2$, then an element
of $C^{\infty}(\Bbb R^N;\Bbb R^N\wedge \Bbb R^N)$ is an
$N\times N$ anti-symmetric matrix valued function mentioned earlier and when
$\ell=N$, it is the determinant.

Notice that 
$$d_\phi \Delta_{\phi}=\Delta_{\phi}d_\phi,\,d_\phi^{\ast} 
\Delta_{\phi}=\Delta_{\phi}d_\phi^{*}\tag 7.8$$ by using (7.6). If we let
$\Delta_{\phi}^{(\ell)}$ be the restriction of  $\Delta_{\phi}$ to forms of degree
$\ell$, we obtain more precisely:
$$d_\phi \Delta_{\phi}^{(\ell)}=\Delta_{\phi}^{(\ell+1)}d_\phi,\,d_\phi^{*} 
\Delta_{\phi}^{(\ell+1)}=\Delta_{\phi}^{(\ell)}d_\phi^{*}.\tag 7.9$$ 

We remark that the above construction is similar to
that of Hodge Laplacian which corresponds  to taking $\phi=0$.
(For a quick overview of the analytical aspects of Hodge theory, see Chap 11.3 in \cite{CFKS}. ~)
We have explicitly
$$-\Delta_\phi^{(0)}={d^*_\phi}{d_\phi}=\sum^N_{j=1} 
z_j^{*}z_j=-\sum^N_{j=1}\frac{\partial^2}{\partial x_j^2}+\Vert 
d\phi\Vert^2-\text {Tr} \text{ Hess }\phi.\tag 7.10$$ For example, if $\phi$ is a
non-degenerate quadratic form, then
$-\Delta_\phi^{(0)}$ is an $N$-dimensional harmonic oscillator. $z_j$  and
$z_j^{*}$ are just the annihilation and creation operators for the harmonic oscillator. More generally, we have
$$\aligned- \Delta_\phi&=\sum\sum z_j 
z_k^{*}dx_j^{\wedge}dx_{k}^{\rfloor}+\sum\sum 
z_k^{*}z_jdx_{k}^{\rfloor}dx_j^{\wedge}\\ &=\sum\sum 
z_k^{*}z_j(dx_j^{\wedge}dx_{k}^{\rfloor}+dx_{k}^{\rfloor}dx_j^{\wedge})+
[z_j,z_k^{*}]dx_j^{\wedge}dx_{k}^{\rfloor}\\
&=\sum z_j^{\ast}z_j+2\sum\sum(\partial_{x_j}\partial_{x_k}\phi) 
dx_j^{\wedge}dx_{k}^{\rfloor}\\ &=-\Delta_\phi^{(0)}\otimes
\Bbb I+2\sum\sum(\partial_{x_j}\partial_{x_k}\phi) 
dx_j^{\wedge}dx_{k}^{\rfloor},\endaligned\tag 7.11$$ where to obtain the third line
from the second, we used (7.5). In particular, with the identification of 1-forms with
$\Bbb R^{N}$ valued functions, we obtain
$$-\Delta_{\phi}^{(1)}=-\Delta_{\phi}^{(0)}\otimes \Bbb I+2\phi''.\tag 7.12$$ 

Since formally $(-\Delta_{\phi}^{(\ell)}u,u)\geq 0$, under appropriate conditions
on $\phi$ at infinity, we can define
$-\Delta_{\phi}^{(\ell)}$ as a self-adjoint operator and that $-\Delta_{\phi}^{(\ell)}$ has compact resolvent, cf. \cite{S1}. 
Moreover $-\Delta_{\phi}^{(\ell)}$ has discrete  
spectrum contained in $[0,\infty)$. The lowest eigenvalue of 
$-\Delta_{\phi}^{(0)}$ is zero and a corresponding eigenfuction is 
$e^{-\phi}$, since this function is annihilated by $d_{\phi}$. This  eigenvalue is
simple, for if $u$ is another eigenfunction associated  to the same eigenvalue, then
$0=(-\Delta_{\phi}^{(0)}u,u)=\Vert  d_{\phi} u\Vert^{2}$ and hence $d_{\phi} u=0$,
which means precisely that $u$ is a multiple of $e^{-\phi}$.

Using (7.8), we obtain the following intertwining property of the spectra:
 $$\sigma (-\Delta_\phi^{(0)})\backslash\{0\}\subset \sigma (-\Delta_\phi^{(1)}).\tag 7.13$$
This is because if $f$ is an eigenfunction of $-\Delta_\phi^{(0)}$:
$$-\Delta_\phi^{(0)}f=\kappa f\tag 7.14$$
with $\kappa>0$, then operating  on (7.14) with $d_\phi$, we obtain
$$-d_\phi \Delta_\phi^{(0)}f=-(d_\phi d^*_\phi) d_\phi f=-\Delta_\phi^{(1)} (d_\phi f)=\kappa d_\phi f.$$
So if $\kappa\neq 0$, then $d_\phi f\neq 0$ is an eigenform for  $-\Delta_\phi^{(1)} $, which is the 
statement in (7.13). 

This is in fact the main reason that we introduced $-\Delta_\phi^{(1)}$.  Using (7.12),
we then obtain that $\sigma (-\Delta_\phi^{(0)})$ has a spectral gap if 
$\phi$ is strictly convex. 
We end this self-adjoint introduction by remarking that if $e^{-\phi}$ is the eigenfunction of the 
Schr\"odinger operator 
$$-\sum \frac{\partial^2}{\partial x_j^2}+V$$
for the lowest eigenvalue $\kappa$, then 
$$-\sum \frac{\partial^2}{\partial x_j^2}+V-\kappa=-\Delta_\phi^{(0)}.$$
From this point of view, the Witten-Hodge Laplacian can be seen as a 
natural generalization of harmonic oscillators when $\phi$ is quadratic.


\Refs\nofrills{References}
\widestnumber\key{CFKSAB}
\ref
\key {\bf Bi}
\by J.-M. Bismut
\paper  The hypoelliptic Laplacian on the cotangent bundle
\jour J. Amer. Math. Soc.
\vol 18
\pages 379-476
\yr 2005
\endref

\ref
\key {\bf Bo1}
\by J. Bourgain
\paper  Periodic nonlinear Schr\"odinger equation and invariant measure
\jour Commun. Math. Phys.
\vol 166
\pages 1-26
\yr 1994
\endref

\ref
\key {\bf Bo2}
\by J. Bourgain
\paper  Invariant measure for the 2D-defocusing nonlinear Schr\"odinger equation
\jour Commun. Math. Phys.
\vol 176
\pages 421-445
\yr 1996
\endref

\ref
\key {\bf Bo3}
\by J. Bourgain
\paper  Nonlinear Schr\"odinger equations
\jour IAS/Park City Mathematics Series
\vol 
\pages 1-157
\yr 1999
\endref


\ref
\key {\bf BS}
\by D. Brydges, G. Slade
\paper  Statistical mechanics of the 2-dimensional focusing nonlinear Schr\"odinger equation
\jour Commun. Math. Phys.
\vol 182
\pages 485-504
\yr 1996
\endref

\ref
\key{\bf CFKS}
\by H. L. Cycon, R. G. Froese, W. Kirsch, B. Simon
\book Schr\"odinger Operators
\publ Springer-Verlag
\yr 1987
\endref

\ref
\key {\bf DD}
\by G. Da Prato, A. Debussche
\paper  Strong solutions to the stochastic quantization equations
\jour Ann. of Prob. 
\vol 31
\pages 1900-1916
\yr 2003
\endref

\ref
\key {\bf EH}
\by J.-P. Eckmann, M. Hairer
\paper  Uniqueness of the invariant measure for a stochastic PDE driven by degenerate noise
\jour Commun. Math. Phys. 
\vol 219
\pages 523-565
\yr 2001
\endref

\ref
\key {\bf GJ}
\by J. Glimm, A. Jaffe
\book Quantum Physics
\publ Springer-Verlag
\yr 1987
\endref

\ref
\key {\bf H}
\by F. Herau
\paper  Short and long time behavior of the Fokker-Planck equation in a confining potential 
and applications
\jour J. Func. Anal.
\vol 244
\pages 95-118
\yr 2007
\endref


\ref
\key {\bf HHS}
\by F. Herau, M. Hitrik, J. Sj\"ostrand 
\paper  Tunnel effect for Kramer-Fokker-Planck type operators
\jour Ann Henri Poincar\'e
\vol 9
\pages 209-274
\yr 2008
\endref

\ref
\key{\bf J}
\by A. Jaffe
\jour Ann Arbor Lecture
\yr 1994
\endref


\ref
\key {\bf KuS}
\by S. Kuksin, A, Shirikyan
\paper  Stochastic dissipative PDE's and Gibbs measure
\jour Commun. Math. Phys.
\vol 213
\pages 291-330
\yr 2000
\endref



\ref
\key {\bf LRS}
\by J. Lebowitz, R. Rose, E. Speer
\paper  Statistical mechanics of the nonlinear Schr\"odinger equation
\jour J. Stat. Phys.
\vol 50
\pages 657-687
\yr 1988
\endref

\ref
\key {\bf McKV}
\by H. McKean, K. Vaninski
\paper  Statistical mechanics of nonlinear wave equations. IV. Cubic Schr\"odinger
\jour Comm. Math. Phys.
\vol 168
\pages 479-491
\yr 1995
\endref


\ref
\key {\bf MCL}
\by Ph. Mounaix, P. Collet, J. Lebowitz
\paper  Nonequilibrium stationary state of a truncated stochastic nonlinear Schr\"odinger equation: Formulation and mean-field approximation
\jour Phys. Rev. E
\vol 81
\pages 
\yr 2010
\endref

\ref
\key {\bf O}
\by C. Odasso
\paper  Ergodicity for the stochastic complex Ginzburg-Landau equations
\jour Annales de IHP, Probabilit\'e et Statistique
\vol 42
\pages 417-454
\yr 2006
\endref


\ref
\key {\bf S1}
\by J. Sj\"ostrand
\paper  Correlation asymptotics and Witten Laplacians
\jour Algebra and Analysis
\vol 8 (1)
\pages 160-191
\yr 1996
\endref

\ref
\key {\bf S2}
\by J. Sj\"ostrand
\paper  Some results on non self-adjoint operators: a survey
\inbook  Further progress in analysis
\publ World Sci. Publ., Hackensack, NJ 
\pages 45-75
\yr 2009
\endref

\ref
\key{\bf Sp}
\by M. Spivak
\book A Comprehensive Introduction to Differential Geometry, Vol. I
\publ Publish or Perish, Berkely
\yr 1970
\endref
 
\ref
\key {\bf TT-NK}
\by J. Tailleur, S. Tanase-Nicola, J. Kurchan
\paper  Kramers equation and supersymmetry
\jour J. Stat. Phys. 
\vol 122(4)
\pages 557-595
\yr 2006
\endref


\ref
\key {\bf W}
\by W.-M. Wang
\paper  Supersymmetry, Witten complex and asymptotics for directional Lyapunov exponents in $\Bbb Z^d$
\jour Ann. Henri Poincare
\yr 2001
\vol 2
\pages No. 2, 237-307
\endref 


\ref
\key {\bf Wi}
\by E. Witten
\paper  Supersymmetry and Morse theory
\jour J. Diff. Geom.
\yr 1982
\vol 17
\pages 661-692
\endref 


\ref
\key {\bf Z}
\by P. Zhidkov
\paper  An invariant measure for the nonlinear Schr\"odinger equation
\jour Dokl. Akad. Nauk SSSR
\vol 317
\yr 1991
\pages No. 3, 543-546
\endref 

\endRefs
\enddocument
\end